\documentclass[vecphys]{svmult}

\usepackage{graphicx}        
\usepackage[bottom]{footmisc}

\usepackage{natbib}
\bibpunct{(}{)}{;}{a}{}{,}
\newcounter{Rco}
\newcommand{\Ionst}[1]{\setcounter{Rco}{#1}\Roman{Rco}}
\newcommand{\Ion}[2]{\mbox{#1\ {\scriptsize\Ionst{#2}}}}
\newcommand{\Ionw}[3]{\mbox{#1\ {\scriptsize\Ionst{#2}}~$\lambda\,#3$\,\AA}}

\newcommand{\loggw}[1]{\mbox{$\log g\hspace{-0.5mm} =\hspace{-0.5mm}  #1$}}

\newcommand{\Teff}{\mbox{$T_\mathrm{eff}$}}
\newcommand{\Teffw}[1]{\mbox{$\Teff\hspace{-0.5mm}=\hspace{-0.5mm}#1\,\mathrm{kK}$}}
\newcommand{\lsv}{\mbox{LS\,V\,$+46^\circ 21$}}
\newcommand{\pnsh}{\mbox{Sh\,2$-$216}}

\newcommand{\APJ}[2]{ApJ, #1, #2 }

\newcommand{\AuA}[2]{A\&A, #1, #2 }
\newcommand{\AuAp}{A\&A, in press}

\newcommand{\MNRAS}[2]{MNRAS, #1, #2 }
\newcommand{\PASP}[2]{PASP, #1, #2 }

\begin{document}

\title*{Spectral Analysis of \\ Central Stars of Planetary Nebulae}
\author{Thomas Rauch\inst{1}\and
Klaus Werner\inst{1}\and
Marc Ziegler\inst{1}\and
Jeffrey W\@. Kruk\inst{2}\and
Cristina M\@. Oliveira\inst{2}}
\authorrunning{Rauch et al\@.} 
\institute{Kepler Center for Astro and Particle Physics,
Institute for Astronomy und Astrophysics,
Eberhard-Karls University, Sand~1, 72076 T\"ubingen, Germany
\texttt{rauch@astro.uni-tuebingen.de}
\and 
Department of Physics and Astronomy, Johns Hopkins University, Baltimore, MD 21218, U.S.A.
\texttt{kruk@pha.jhu.edu}}
%
%
\maketitle

\setcounter{footnote}{0}

\begin{abstract}
Spectral analysis by means of NLTE model atmospheres has presently arrived at a high level of sophistication. 
High-resolution spectra of central stars of planetary nebulae can be reproduced in detail from the infrared 
to the X-ray wavelength range.
\
In the case of \lsv, the exciting star of \pnsh, we demonstrate the state-of-the-art in the determination 
of photospheric properties like, e.g.,  effective temperature (\Teff), surface gravity ($g$), and abundances of 
elements from hydrogen to nickel. 
From such detailed model atmospheres, we can reliably predict the ionizing spectrum of a central star which is a 
necessary input for the precise analysis of its ambient nebula.

NLTE model-atmosphere spectra, however, are not only accessible for specialists. 
In the framework of the \emph{German Astrophysical Virtual Observatory} (\emph{GAVO}), we provide pre-calculated
grids of tables with synthetic spectra of hot, compact stars as well as a tool to calculate individual 
model-atmosphere spectra in order to make the use of synthetic stellar spectra as easy as the use of
blackbody flux distributions had been in the last century.
\end{abstract}

\section{Introduction}
\label{sectintroduction}
A reliable determination of properties of planetary nebulae (PNe) requires precise
knowledge about their central stars (CS).
A photoionization code may be perfect, yet still provide inaccurate results if 
the model spectrum of the exciting star does not match the actual spectrum of the star.

\begin{figure}[ht]
\centering
\includegraphics[width=\textwidth]{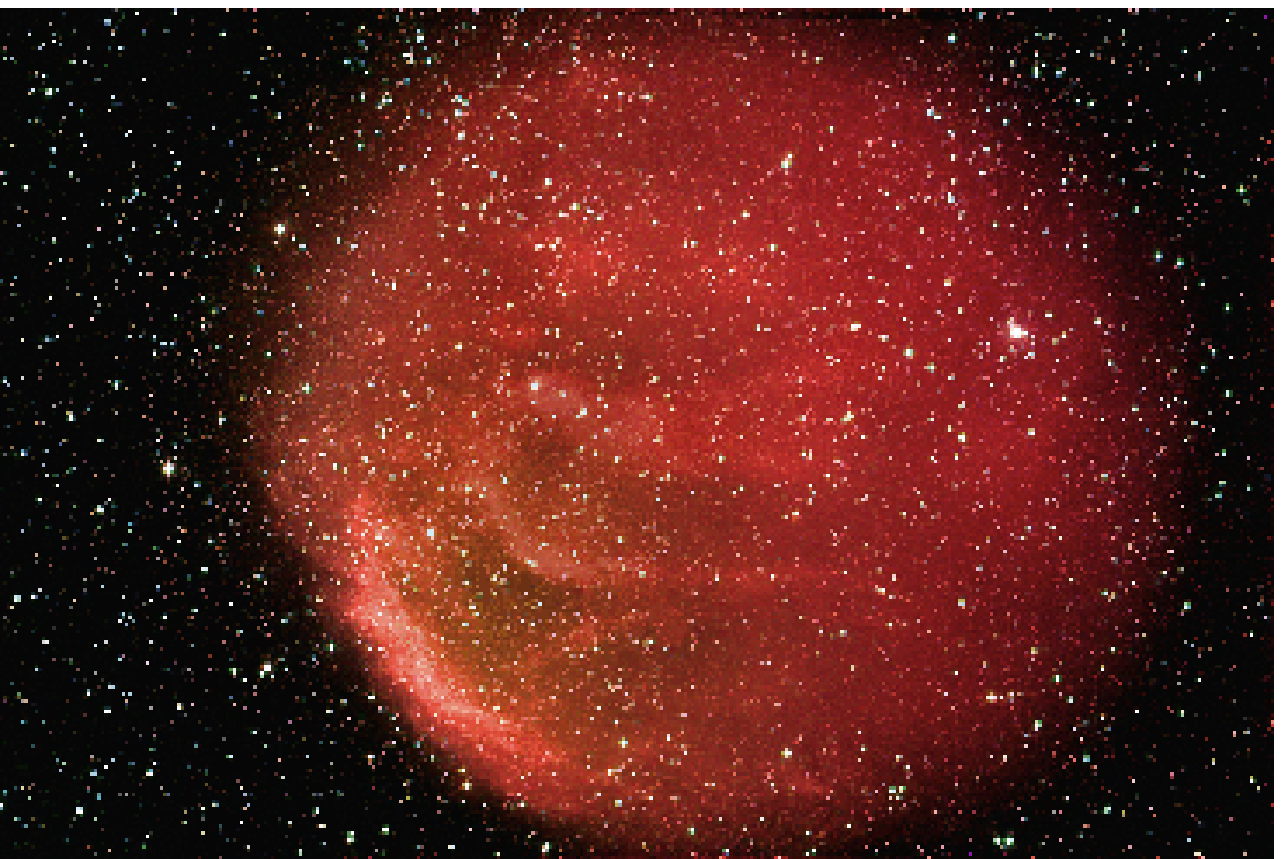}
\caption{Composite wide-field image (North is up and East is left)
of \pnsh\ (apparent diameter of 100') taken by Dean Salman (http://www.deansalman.com).
The exposure time is 600\,min in H\,$\alpha$ (red) and 180\,min in [O\,III] (green).
Note that \pnsh\ has an interaction with the interstellar medium \citep*[ISM;][]{tmn1995}, its CS left the geometrical
center about 45\,000 years ago \citep*{kea2004} and is presently located nearly halfway towards the eastern 
rim \citep{cr1985}.}
\label{fig:sh2216}
\end{figure}

In the last two decades both, observational techniques as well as numerical methods in
theory have been strongly improved. State-of-the-art NLTE model-atmosphere codes like, e.g.,
\emph{TMAP}\footnote{http://astro.uni-tuebingen.de/\raisebox{.2em}{\small $\sim$}rauch/TMAP/TMAP.html}, 
the \emph{T\"ubingen NLTE Model Atmosphere Package} \citep{wea2003, rd2003},
calculate plane-parallel, chemically homogeneous models in hydrostatic and radiative equilibrium
which consider opacities of all elements from hydrogen up to the iron-group \citep{r1997, r2003} 
and thus, are well suited to provide synthetic ionizing spectra for hot, compact stars.

In this paper, we use \lsv, the central star of \pnsh\ (Fig\@. \ref{fig:sh2216}),  
in order to demonstrate the capabilities of
\emph{TMAP} to reproduce the UV spectra of hot stars (Sect\@. \ref{sect:spectral}).

The perspectives of spectral analysis in the framework of the \emph{Virtual Observatory} 
(\emph{VO}) are described by the example of
synthetic spectra calculated by \emph{TMAP} (Sect\@. \ref{sect:VO}).

\section{Spectral analysis of hot, compact stars}
\label{sect:spectral}
Stars with high \Teff\ (in the case of CS up to about 200\,kK)
have their flux maximum in the EUV. Since precise NLTE
spectral analysis needs metal lines (of highly ionized species) in order to determine
\Teff\ (evaluation of ionisation equilibria) and elemental abundances, high
signal-to-noise (S/N) and high-resolution UV spectra are necessary. These were
provided by instruments aboard the HST\footnote{Hubble Space Telescope}, namely
FOS\footnote{Faint Object Spectrograph} 
(working $1990 - 1997$, wavelength range $1150\,\mathrm{\AA} < \lambda < 8000\,\mathrm{\AA}$, resolution $\approx 1.9$\,\AA),
GHRS\footnote{Goddard High-Resolution Spectrograph} 
($1990 - 1997$, $1150\,\mathrm{\AA} < \lambda <  3000\,\mathrm{\AA}$, resolving power $R \le 80\,000$),
STIS\footnote{Space Telescope Imaging Spectrograph} 
($1997 - 2004$, $1150\,\mathrm{\AA} < \lambda <  3175\,\mathrm{\AA}$, $R \le 114\,000$)
and by
FUSE\footnote{Far Ultraviolet Spectroscopic Explorer} 
($1999 - 2007$, $904\,\mathrm{\AA} < \lambda <  1190\,\mathrm{\AA}$, $R \le 20\,000$).

\begin{figure}[ht]
\centering
\includegraphics[width=\textwidth]{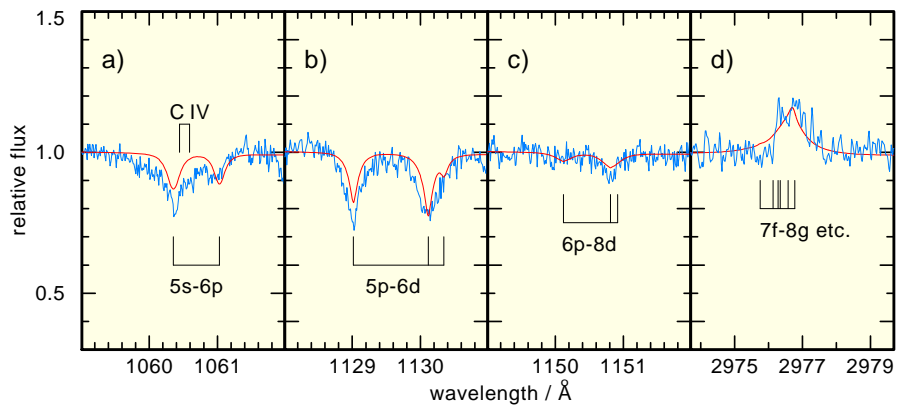}
\caption{Identification of \Ion{Ne}{8} lines in the FUSE (panels a -- c) and HST/GHRS (panel d)
         observations of the  extremely hot (\Teffw{170}) CS of 
         MWP\,1 \citep*[PN\,G080.3$-$10.4,][]{mwp1993}, 
         RX\,J2117.1+3412 \citep{wea2007b}. Note that the \Ion{C}{4} doublet in panel a is not
         included in the model.}
\label{fig:neviii}
\end{figure}

The photospheric spectra of CS are characterized by a few, broad and shallow, absorption lines from
highly ionized species like, e.g., 
\Ion{He}{2}, 
\Ion{C}{4}, 
\Ion{O}{6}, 
\Ion{Ne}{7} \citep*{wea2004}, 
\Ion{Ne}{8} \citep*{wea2007b}, 
\Ion{Si}{4}, 
\Ion{Si}{5} \citep*{jea2007}, 
\Ion{Si}{6} \citep*{jea2007}, 
\Ion{S}{6}  \citep{mea2002}, 
\Ion{Ar}{6} \citep*{rea2007}, 
\Ion{Ar}{7} \citep*{wea2007a}.
As an example, in Fig.\,\ref{fig:neviii} we show recently identified \Ion{Ne}{8} lines in the spectrum the PG\,1159-type CS
RX\,J2117.1+3412.

In a recent spectral analysis of \lsv\ by means of \emph{TMAP}
NLTE model atmospheres, \citet{rea2007} 
considered the elements
H, He, C, N, O, F, Mg, Si, P, S, Ar, Ca, Sc, Ti, V, Cr, Mn, Fe, Co, and Ni.
H\,$-$\,Ar were represented by ``classical'' model atoms \citep{r1997} partly taken from 
\emph{TMAD}\footnote{http://astro.uni-tuebingen.de/\raisebox{.2em}{\small $\sim$}rauch/TMAD/TMAD.html}, 
the \emph{T\"ubingen Model Atom Database}.
For Ca$-$Ni individual model atoms are constructed by
\emph{IrOnIc} \citep*{rd2003}, using a statistical approach in order to treat the
extremely large number of atomic levels and line transitions by the introduction
of ``super-levels'' and ``super-lines''. In total 686 levels are treated in NLTE,
combined with 2417 individual lines and about 9 million iron-group lines, 
taken from \citet{k1996} as well as from the OPACITY and IRON projects \citep{sea1994,hea1993}
\citep[see][and references therein for details]{rea2007}.

More than 2000 spectral lines could be identified in the FUSE and STIS spectra of \lsv. This is 
about 95\% of all spectral features in this wavelength range. It has been possible to identify lines of, e.g., 
\Ion{Si}{5} (for the first time in the spectrum of this star),
\Ion{Mg}{4} (for the first time in such objects), and 
\Ion{Ar}{6} (for the first time in any star).
However, it is likely that the still unidentified lines simply stem from the most prominent ions as well, 
but their wavelengths are not precisely known because no laboratory measurements exist.
E.g., \citet{k1996} lists fewer than about one percent of his \Ion{Fe}{6} and \Ion{Fe}{7}
lines as having measured wavelengths.  

\begin{figure}[ht]
\centering
\includegraphics[width=\textwidth]{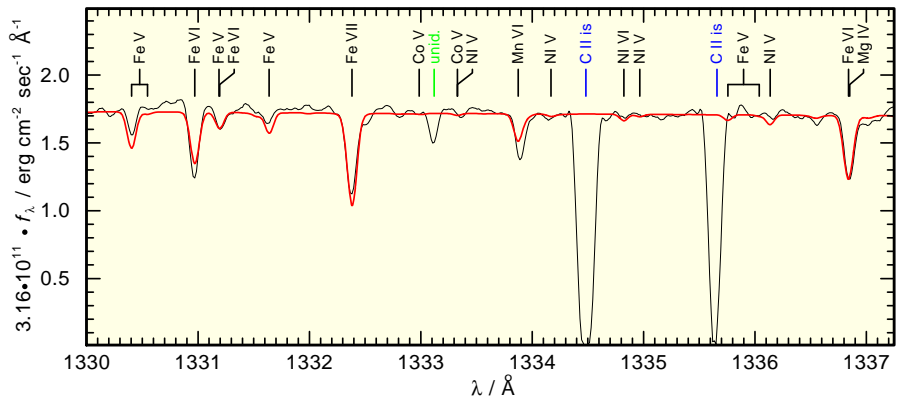}
\caption{Modelling the ionisation equilibrium of \Ion{Fe}{5} $-$ \Ion{Fe}{7}. It is well matched at \Teffw{95}.
Ionisation equilibria are sensitive indicators for \Teff. 
The marks indicate the positions of identified lines (is denotes interstellar lines, unid\@. is an unidentified
line).}
\label{fig:fe}
\end{figure}

The evaluation of ionisation equilibria of different elements and ionisation stages (Fig\@. \ref{fig:fe})
allowed to determine \Teffw{95\pm 2} (and \loggw{6.9\pm 0.3}) 
with high accuracy \citep[cf\@.][for further details]{rea2007}. 

\begin{figure}[ht]
\centering
\includegraphics[width=\textwidth]{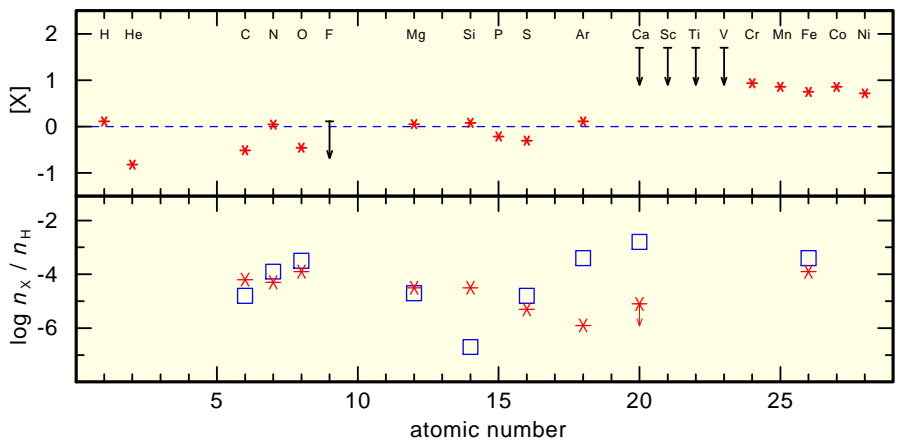}
\caption{Top: Photospheric element abundances of \lsv\ determined from detailed line profile fits. 
              [x] denotes log (mass fraction / solar mass fraction) of species x.
              The dashed, horizontal line indicates the solar abundance values \citep*{ags2005}.
              For F, Ca, Sc, Ti, and V upper limits can be found only.
         Bottom: Comparison of the elemental  number ratios found in our spectral analysis (red stars)
              compared to predictions (blue squares) of diffusion calculations for DA models \citep{cea1995} 
              with \Teffw{95} and \loggw{7}.
}
\label{fig:X}
\end{figure}

The abundance pattern found in \lsv\ (Fig\@. \ref{fig:X}) is the result of the interplay of gravitational settling
(in the case of lighter elements, like He and C) and radiative levitation (for the iron-group elements).
The results of \citet{rea2007} are mostly in good agreement with diffusion models for DA white dwarfs 
\citep[Fig\@. \ref{fig:X}]{cea1995}. Unfortunately, it is not possible to compare the results for all 
elements from H $-$ Ni. One reason for this is the lack of reliable atomic data for all the species
and especially their higher ionisation stages. Every endeavor should be made to improve the atomic data, 
as the gaps in our knowledge in this area hamper reliable modeling in many fields of astronomy, 
not only in stellar spectral analysis.

Another problem for the precise analysis of UV spectra of hot stars is the contamination by
interstellar absorption. \citet{rea2007} have used the \emph{OWENS} program in order to model
simultaneously both, the stellar as well as the interstellar absorption lines
\citep[cf\@.][for more details]{oea2007}.  

\begin{figure}[ht]
\centering
\includegraphics[width=\textwidth]{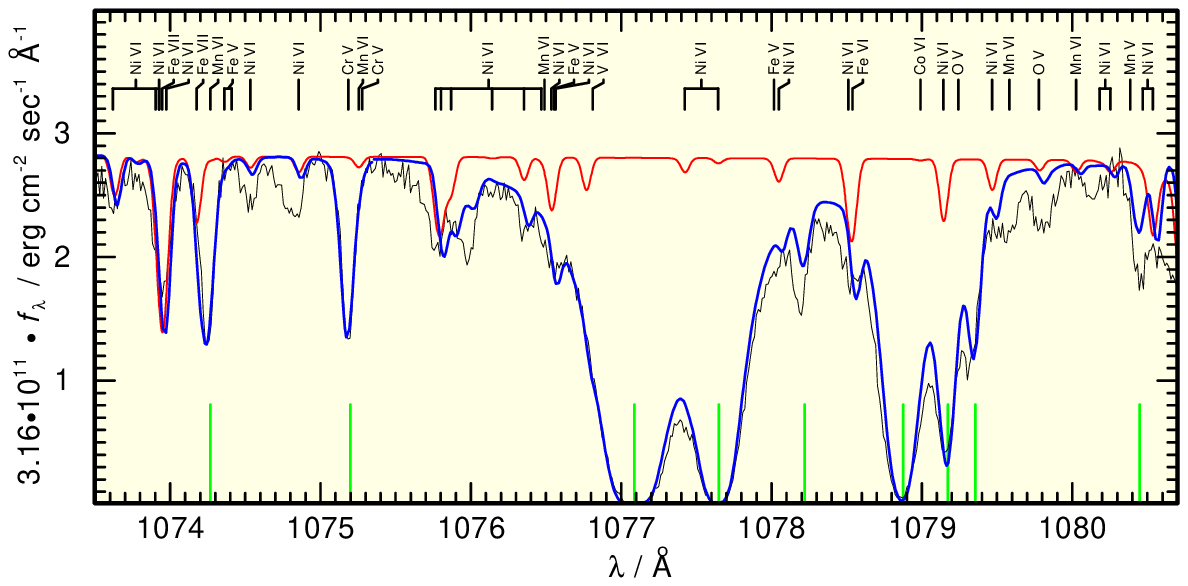}
\caption{Modelling of the ISM absorption lines with \emph{OWENS}.
         Section of the FUSE spectrum of \lsv\ compared 
         with the final model of \citet[][thin line]{rea2007} and
         with the combined ISM + model-atmosphere spectrum (thick line).
         Identified lines are marked at top.
         Most of the interstellar absorption is due to H$_2$ (the line positions
         are marked by vertical bars at the bottom).
         The synthetic spectra are normalized to match the local continuum. 
         Note that it is possible to  unambiguously identify a few isolated photospheric lines, 
         e.g\@. \Ionw{Fe}{7}{1073.9}, which are suitable for spectral analysis.}
\label{fig:owens}
\end{figure}

The spectral analysis of \lsv\ \citep{rea2007} has made use of the most detailed model atmospheres which
were calculated with \emph{TMAP} so far. The interested reader may have a look at the large online
plots\footnote{http://vizier.cfa.harvard.edu/viz-bin/ftp-index?J/A\%2BA/470/317} 
which are provided by \citet{rea2007} in order to convince oneself that the agreement between
observation and theory is impressive.

\section{TMAP and the Virtual Observatory}
\label{sect:VO}

In the framework of the \emph{German Astrophysical Virtual Observatory} project 
(\emph{GAVO}, 
please note that the URLs given below will change to the \emph{GAVO} portal\footnote{http://www.g-vo.org/portal/} later),
we use \emph{TMAP} to provide synthetic model-atmosphere spectra.
For the \emph{VO} user, there are three access levels:

\begin{itemize}
\item[$\bullet$]
The most easy way is to use pre-calculated grids of model-atmosphere flux tables
(\emph{TMAF}\footnote{http://astro.uni-tuebingen.de/\raisebox{.2em}{\tiny $\sim$}rauch/TMAF/TMAF.html}).
\item[$\bullet$]
A WWW interface allows to calculate simple (H+He+C+N+O) model atmospheres based on pre-defined model atoms 
for an individual object
without profound experience with \emph{TMAP}
(\emph{TMAW}\footnote{http://astro.uni-tuebingen.de/\raisebox{.2em}{\tiny $\sim$}rauch/TMAW/TMAW.html}).
\item[$\bullet$]
For a more detailed analysis, the \emph{VO} user may use
atomic data which are 
provided within the \emph{T\"ubingen Model-Atom Database}
\emph{TMAD}\footnote{http://astro.uni-tuebingen.de/\raisebox{.2em}{\tiny $\sim$}rauch/TMAD/TMAD.html}
in order to construct a custom model atom (H $-$ Ni)
which is suited for the investigation on a particular star.
\end{itemize}

With this approach, a \emph{VO} user may compare observation and synthetic spectra at three stages:
The easiest and fastest way is the inter- or extrapolation within the
flux-table grids. For a more detailed analysis, the \emph{VO} user may improve the 
fit to the observation by the calculation of adjusted model atmospheres with individual
stellar photospheric parameters via \emph{TMAW}. A more experienced \emph{VO} user may tailor own 
atomic-data files for an individual analysis and then calculate model atmospheres and flux tables 
with these.

\emph{TMAP} flux tables are already incorporated into the photoionization codes 
\emph{CLOUDY} \citep{fea1998} and \emph{\mbox{MOCASSIN}} \citep[e.g\@.][]{eea2003, eea2005}. 
We will create a WWW interface for the control of the 3D-code \emph{\mbox{MOCASSIN}}
which makes then directly use of model-atmosphere fluxes within the \emph{GAVO} database.
However, any photoionization code may benefit from the synthetic spectra provided by the \emph{VO}.

\section{Conclusions}
\label{sect:conclusion}

NLTE model atmospheres which are calculated with state-of-the-art codes like, e.g., \emph{TMAP}
have presently arrived at a high level of sophistication and are successfully employed for
the analysis of, e.g., central stars of planetary nebulae.

The spectral analysis of high-resolution and high-S/N UV spectra has shown that the available
atomic data is not complete and partly not accurate enough even for the most abundant species \citep[cf\@.][]{rea2007}

Atomic physicists are therefore challenged to measure atomic data precisely for many elements and
for higher ionization stages.
Better atomic and, e.g., line-broadening data
will then strongly improve future spectral analyses and thus, make determinations of
photospheric properties more reliable. 

NLTE model-atmosphere fluxes are now easy to access and it is recommended to use them instead of
blackbody flux distributions which are only a bad approximation of a real star.

The \emph{VO} has become an invaluable source for the future spectral analysis.
Any astronomer who has data, either observations or models, should be aware of this
and contribute to the improvement of databases and tools within the \emph{VO}.
\acknowledgement
T.R\@. is supported by the \emph{German Astrophysical
Virtual Observatory} project of the German Federal Ministry of Education
and Research under grant 05\,AC6VTB. 
J.W.K\@. is supported by the FUSE project, funded by NASA contract NAS532985.
This research has made use of the SIMBAD Astronomical Database, operated at CDS, Strasbourg, France.
This work has been done using the profile-fitting procedure \emph{OWENS} developed by M\@. Lemoine and the FUSE French Team.
%
%



\end{document}